\documentclass[10pt,letterpaper]{article}
\usepackage[top=0.85in,left=1.6in,footskip=0.75in]{geometry}

\usepackage{amsmath,amssymb}

\usepackage{changepage}

\usepackage[utf8x]{inputenc}

\usepackage{textcomp,marvosym}

\usepackage{cite}

\usepackage{nameref,hyperref}

\usepackage[right]{lineno}

\usepackage{microtype}
\DisableLigatures[f]{encoding = *, family = * }

\usepackage[table]{xcolor}

\usepackage{array}

\usepackage{soul}

\usepackage{caption,tabularx,booktabs}
\usepackage{pdfpages}

\newcolumntype{+}{!{\vrule width 2pt}}

\newlength\savedwidth

\usepackage[aboveskip=1pt,labelfont=bf,font=small,labelsep=period,singlelinecheck=off]{caption}

\makeatletter
\renewcommand{\@biblabel}[1]{\quad#1.}
\makeatother

\usepackage{lastpage,fancyhdr,graphicx}
\usepackage{epstopdf}
\usepackage{dblfloatfix}
\usepackage{xcolor} 
\usepackage{subcaption}	
\usepackage{multicol}
\usepackage{graphicx}	
\usepackage{mwe}
\usepackage{textcomp}
 \usepackage{steinmetz}

\usepackage[title]{appendix}
\usepackage{algorithm,algpseudocode}
\newcounter{algsubstate}

\usepackage{cite}

\usepackage{ifthen}
\usepackage{amsmath}
\usepackage{amssymb}
\usepackage{theorem}
\usepackage{xspace}
\usepackage{calc}
\usepackage{amsfonts}
\usepackage{multicol}
\usepackage{graphicx}	
\usepackage{mwe}
\usepackage{textcomp}
 \usepackage{steinmetz}
 \usepackage{xcolor}

\newcommand{\artworks}{0}
\newcommand{\figures}{1}

\usepackage{ifthen}

\newcommand{\trytoturnpage}{\vspace*{20em}\par\noindent}
\newcommand{\myfig}[1]{\ifthenelse{\artworks=1}{\begin{figure}[f]\trytoturnpage}{\begin{figure}[#1]}}
\newcommand{\mytab}[1]{\ifthenelse{\artworks=1}{\begin{table}[f]\trytoturnpage}{\begin{table}[#1]}}
\newcommand{\myfigstar}[1]{\ifthenelse{\artworks=1}{\begin{figure*}[f]\trytoturnpage}{\begin{figure*}[#1]}}
\newcommand{\mytabstar}[1]{\ifthenelse{\artworks=1}{\begin{table*}[f]\trytoturnpage}{\begin{table*}[#1]}}

\newcommand{\mycaption}[1]{\ifthenelse{\artworks=1}{\vspace*{10em}\caption{#1}}{\caption{#1}}}

\newcommand{\myfigend}{\ifthenelse{\artworks=1}{\trytoturnpage\end{figure}}{\end{figure}}}
\newcommand{\myfigstarend}{\ifthenelse{\artworks=1}{\trytoturnpage\end{figure*}}{\end{figure*}}}

\newcommand{\mytabend}{\ifthenelse{\artworks=1}{\trytoturnpage\end{table}}{\end{table}}}
\newcommand{\mytabstarend}{\ifthenelse{\artworks=1}{\trytoturnpage\end{table*}}{\end{table*}}}

\newcommand{\mycenterwmf}[3]{\ifthenelse{\figures=1}{\centerwmf{#1}{#2}{#3}}{\vskip#2\medskip}}
\newcommand{\myspecial}[1]{\ifthenelse{\figures=1}{\special{#1}}{}}
\newcommand{\mycentereps}[3]{\ifthenelse{\figures=1}{\centereps{#1}{#2}{#3}}{\vskip#2\medskip}}

\newsavebox{\fminibox}
\newlength{\fminilength}
\newenvironment{fminipage}[1][\linewidth]
	{ \setlength{\fminilength}{#1}\addtolength{\fminilength}{-2\fboxsep}%
					       \addtolength{\fminilength}{-2\fboxrule}%
	   \begin{lrbox}{\fminibox}\begin{minipage}{\fminilength}}
	{ \end{minipage}\end{lrbox}\noindent\fbox{\usebox{\fminibox}}}

\newcommand{\gComment}[1]{}

{
\theoremheaderfont{\normalfont\scshape}

}

\usepackage{lipsum}

\newcommand{\begq}{\begin{equation}}
\newcommand{\eeq}{\end{equation}}

\begin{document}
\vspace*{0.2in}

\begin{flushleft}

\newcommand\blfootnote[1]{%
  \begingroup
  \renewcommand\thefootnote{}\footnote{#1}%
  \addtocounter{footnote}{-1}%
  \endgroup
}

{\Large
\textbf\newline{\textbf{Introducing the modularity graph: \\an application to brain functional networks}}
}

\bigskip

Tiziana Cattai\textsuperscript{1},
Camilla Caporali\textsuperscript{1},
Marie-Constance Corsi\textsuperscript{2},
Stefania Colonnese\textsuperscript{1}
\\
\bigskip
\textbf{1} Dept. of Information Engineering, Electronics and Telecommunication, Sapienza University of Rome, Italy 
\\
\textbf{2} Sorbonne Université, Institut du Cerveau – Paris Brain Institute -ICM, CNRS\\
Inria, Inserm, AP-HP, Hopital de la Pitié Salpêtrière, F-75013, Paris, France
\\
\bigskip

Corresponding author: tiziana.cattai@uniroma1.it

\blfootnote{This work has been submitted to the IEEE for possible publication. Copyright may be transferred without notice, after which this version may no longer be accessible}

\blfootnote{We acknowledge financial support under the National Recovery and Resilience Plan (NRRP), Mission 4, Component 2, investment 1.1, Call for tender No. 104 published on 2.2.2022 by the Italian Ministry of University and Research (MUR), funded by the European Union – NextGenerationEU– Project Title  QT-SEED Quality-of-life Technological and Societal Exploitation of ECG Diagnostics   – CUP  B53D23002460006  - Grant Assignment Decree No. D.D. 960   adopted on  30/06/2023 by the Italian Ministry of  University and Research (MUR)}

\end{flushleft}
\bigskip

\begin{abstract} 
In  signal processing, exploring complex systems through network representations has become an area of growing interest. This study introduces the modularity graph, a new graph-based feature, to highlight the relationship across the graph communities. After showing an application to the random graph class known as Stochastic Block Model, we consider the brain functional connectivity network estimated from real EEG data. The modularity graph provides a quantitative framework for examining the interactions between neuron clusters within the brain's network. The modularity graph works alongside multiscale community detection algorithms, thereby enabling the identification of community structures at various scales. 

After introducing the modularity graph,  we apply it to the brain functional connectivity network, estimated from publicly available EEG recordings of motor imagery experiments.  Statistical  analysis across multiple scales shows that  the modularity graph differs for the distinct   brain connectivity states associated with various motor imagery tasks.

This work emphasizes the application of signal on graph processing techniques to understand brain behavior during specific cognitive tasks, leveraging the novel modularity graph to  identify patterns of brain connectivity in different cognitive conditions. This approach sets the stage for further signal on graph analysis to devise brain network modularity, and to gain  insights into the   motor imagery mechanisms.
\end{abstract}

\section{Introduction}
Recent signal processing literature witnesses an increasing interest in the analysis of graphs originating from complex, noisy signals. Stemming on cutting-edge signal on graph processing techniques, such as multiscale community detection, herein  we propose a novel approach to represent the modularity of a graph,
and 
we apply it to brain networks, to highlight the correlations between neuronal signals
observed through 
brain raw EEG data acquired during 
 cognitive tasks such as motor imagery.

In the literature,
the human brain is studied as a complex network, where neurons interact during cognitive and motor tasks\cite{rubinov2010complex}. The graph node identification is usually related to the neuroimaging technique used to record brain data, while the links are typically defined using functional connectivity, which statistically measures the interactions between nodes. 
The brain network is  organized  in groups of nodes or communities \cite{sporns2016modular}, and its community structure is crucial  in understanding the network functionality \cite{bassett_dynamic_2011,bassett_robust_2013,betzel_multi-scale_2017} and identifying characteristics of brain disease states, such as Alzheimer or Parkinson disease \cite{chen2001patterns,calderone2016comparing}.

\newcommand{\mysep}{\phantom{xxxXxxxxxxxxxxxxxxxxxx}}
 \begin{figure*}[h!]
   \centering{\includegraphics[scale=0.6]{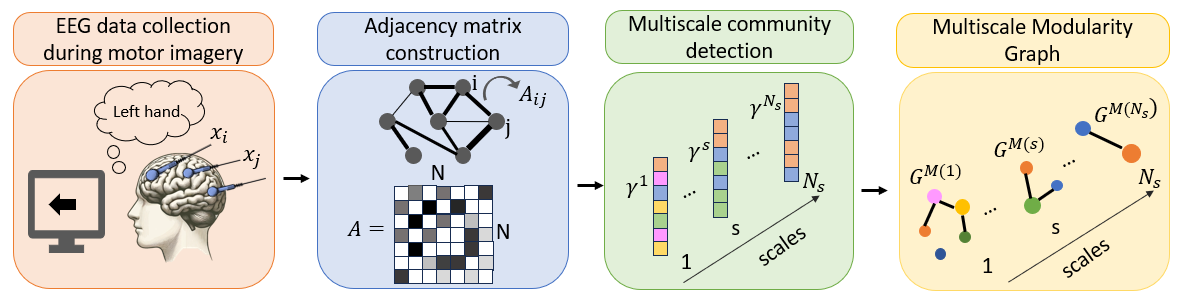}}
\\a) \mysep b) \mysep c) \mysep d)
    \caption{Visual abstract of the proposed approach. In synthesis, we firstly collect the signal measurements on the graph nodes, i.e. on EEG electrodes in this case, then we build the adjacency matrix by evaluating imaginary coherence from each possible couple of node signals. Then, we apply the multiscale community detection algorithm and we obtain a community label vector that assigns a community label to each graph node for each scale. Finally, we construct the modularity graphs at each scale to have a cluster characterization during each mental state. }
    \label{fig:abs}
\end{figure*}

Herein, we propose a novel high-level representation of the modularity, namely the modularity graph, and show its application to the graph communities of the human brain during complex cognitive tasks, such as motor imagery. 
Comparing the modularity graph approach with existing methods \cite{candia2024measures} that describe the cluster presence in the brain, we present a more structured way to characterize communities. Indeed, while the importance of brain graph communities is nowadays well established, their characterization is still challenging.
The research on signal processing has  developed  several  methods to identify graph communities \cite{zhou2019analysis}. Those methods  detect the graph clusters based on the properties of their Laplacian eigenvectors, or on modularity optimization algorithms \cite{blondel2008fast} or leveraging novel graph signal processing (GSP) tools \cite{ortega2018graph}. GSP has been already explored in the context of brain networks \cite{huang2018graph}, providing useful tools for graph modeling \cite{cattai2022eeg} and denoising \cite{cattai2021improving}. The application of GSP to community detection led to algorithms accounting for the correlated nature of signals   lying at graph nodes. In \cite{tremblay2014graph} a method based on graph wavelet transform has been introduced to identify communities at various scales by estimating correlations between wavelets centered on different nodes, therefore revealing node similarities for clustering.

The proposed modularity graph characterizes the communities, obtained at the output of   the graph inference and the  multiscale communities detection tasks. At each scale, the modularity graph nodes represent the detected communities, while the links represent the 
interactions between nodes belonging to the different communities. The modularity graph is independent of the community labels order, allowing seamless comparison of the modularity observed in different experiments, e.g., in case of brain signals, of different subjects.
\\
In the following, we apply the modularity graph to the class of Stochastic Block Model graphs,  observed in the presence of additive noise. Then,  we apply it to real EEG data collected in a public dataset modularity graph recorded during motor-imagery experiments. The statistical analysis of multiscale modularity graphs can be leveraged to discriminate between different mental states. 
Without loss of generality,  Fig. \ref{fig:abs}  summarizes the application  of the proposed approach to the case of GSP of raw EEG signals.

The structure of the paper is as follows: after formally introducing the novel modularity graph in Sec.\ref{sec:multmodgraph}, in Sec.\ref{sec:sbm} we apply it to the relevant graph family of the Stochastic Block Models; then, in Sec.\ref{sec:eeg} we consider graphs built on real EEG signals and we show that the modularity graph statistically differs in different brain cognitive states, corresponding to motion imagery and rest tasks.

\section{Multiscale Modularity Graph}
\label{sec:multmodgraph}
The  human brain is often modeled as a graph \cite{barabasi2013network}. The graph $\mathcal{G} = (\mathcal{V},\mathcal{E})$ is constituted by a set of $N$ vertices ( or nodes)  connected by the edges $\mathcal{E}$ (or links). 
In case of EEG based applications, the EEG signals $\mathbf{x}_i$ are collected at each electrode $i$, as exemplified in Fig.\ref{fig:abs}a); the electrode corresponds to the $i$-th node of the graph $\mathcal{G}$. Then, 
the adjacency matrix $A$ (see Fig.\ref{fig:abs}b) ) is built by leveraging a temporal or spectral similarity metric $\mathcal{S}(\cdot,\cdot)$ between the signals acquired at the node pair $(i,j)$
\begin{equation}
    A_{ij} = \mathcal{S}\left(\mathbf{x}_i, \mathbf{x}_j\right).
    \label{eq:band}
\end{equation}
The similarity measure $\mathcal{S}\left(\mathbf{x}_i, \mathbf{x}_j\right)$ is typically related to the estimated  power spectra $\hat{P}_{x_i}(\omega)$, $\hat{P}_{x_j}(\omega)$ of the  signals ${x}_i$ and ${x}_j$, and to their estimated cross-spectrum  $\hat{P}_{x_ix_j}(\omega)$.
Among others, a possible choice of the similarity metric is written as the following average over a set $\mathcal{B}$ of frequency intervals
$$\mathcal{S}\left(\mathbf{x}_i, \mathbf{x}_i\right)=\frac{\sum_{\omega\in\mathcal{B} }IC_{ij}(\omega)}{\left|\left| \mathcal{B} \right|\right|} \label{eq:b}$$ where $IC_{ij}(\omega)$ denotes the imaginary coherence, defined as:
\begin{equation}
    IC_{ij}(\omega) = \frac{\left|{\Im(\hat{P}_{ij}(\omega))}\right|}{\sqrt{\hat{P}_i(\omega)\cdot\hat{P}_j(\omega)}}
    \label{eq:icoh}
\end{equation}

With these positions, the adjacency matrix $A \in \Re^{NxN}$ contains the information of the node interactions: the $ij$-th element $A_{ij}$  is equal to the weight of the connection between nodes $i$ and $j$ and zero when there is no  edge between node i and j. Therefore, it reflects the  neuroscience definition of the graph edge as a measure of functional connectivity (FC)  \cite{gonzalez2021network}.  Its estimate is challenging and strongly application-dependent. 

Herein, we seek a high-level description of the connectivity, representing the similarity between  groups of graph nodes.
To this aim,  a key graph property is the modularity, i.e. the organization into clusters.
If the graph $\mathcal{G}$ has a community structure, the node set $\mathcal{V}$  can be partitioned into $P$ communities, and the $i$-th node is equipped with a community label $\gamma_i\in\{0,...P-1\}$. 
In   assortative graphs, 
 nodes in the same community have a link with higher probability with respect to nodes in different communities \cite{newman2002assortative}.

Following this model, the generic element of the adjacency matrix $A_{ij}$ of an assortative graph is derived following :
\begin{equation} \begin{split}
&p_{A_{ij}|\gamma_i,\gamma_j}(A_{ij}|\gamma_i,\gamma_j)=\\
&\left\{
\begin{array}{cc}
   p_1
   \delta (A_{ij})+(1- p_1
   ) \delta (A_{ij}-1)
    & \gamma_i=\gamma_j \\
   p_0
   \delta (A_{ij})+(1- 
   p_0
   ) \delta (A_{ij}-1)   & \gamma_i\neq\gamma_j
\end{array}
\right.  
\end{split}
\label{eq:probmodel}
\end{equation}
where  $ p_1
$ and $p_0
$ are the probabilities to have  an edge between  vertices belonging to the same or different communities respectively, with $p_1>p_2$.
The graph communities are typically represented by means of the binary affinity matrix $\boldsymbol{\Gamma} \in \Re^{N \times P}$ whose  $p$-th column  with $i$-th coefficient equal to $1$ iff the $i$-th node is included in the  $p$-th community. 

For assortative graphs, signal processing provides relevant solutions to identify the communities,  integrating tools as spectral wavelet transform and graph network theory, and the community detection algorithms  account for the properties of the signals defined on a networked domain \cite{ortega2018graph}.
In particular, we resort here to a state-of-the-art method to identify communities \cite{tremblay2014graph}. It leverages the spectral graph wavelet transform \cite{hammond2011wavelets}, by providing a multiscale community structure, particularly suited to graphs built on real signals where the number of communities is not a priori known. The method in \cite{tremblay2014graph} leverages the local information
naturally contained in graph wavelets, and it derives the clusters from correlation measures of wavelets related to different nodes.

By following this approach, at the different scales $s$, $s=0,\cdots N_s-1$, we obtain 
a scale-dependent community vector label $\mathcal{\gamma}^{s}\in \{0,1,\cdots P-1\}$ for each trial, i.e. repetition of the same experiments or, more generically, observations of the same graph. Fig.\ref{fig:abs}c) exemplifies the set of community vectors at different scales.
It is important to underline that the numerical value of the element  $\gamma^{s}_i$ is dependent on the labeling order selected by the community detection algorithm. The community labeling can change across different trials, even if the community structure does not change, making it difficult to study the clusters.
With these positions, the graph is represented by a set of binary  affinity matrices at different scales $$\boldsymbol{\Gamma}^{(s)},\;s=0,\cdots N_s-1$$
where the  $p$-th column  of each affinity matrix $\boldsymbol{\Gamma}^{(s)} \in \{0,1\}^{N \times P_s}$ has the $i$-th coefficient equal to $1$ iff the $i$-th node  belongs to the  $p$-th community, and where the number of communities $P_s$ varies across the scales.
Let us now introduce our novel multiscale modularity graph $\mathcal{G}^{\mathcal{M}\;(s) }$,  $s=0,\cdots N_s-1$, which models the interactions inter and intra communities at scale $s$, capturing the fundamental graph structure characteristics.

The    $s$-th scale adjacency matrix  $A^{\mathcal{M}\;(s)}_{pq}$ of the modularity graph $\mathcal{G}^{\mathcal{M}\;(s) }$ has $P\times P$ elements, defined as:

\begin{equation}
A^{\mathcal{M}\;(s)}_{pq}=\sum\limits_{i=0}^{N-1}\sum\limits_{j=0}^{N-1}\delta(\gamma_i^{(s)},p)\delta(\gamma_j^{(s)},q)\cdot A_{ij}
\label{eq:AM}
\end{equation}

where $p,q\!=\!0,\cdots P\!-\!1$ and $\delta(u,v)$ denotes the  binary indicator function 
\begin{equation}
\delta(u,v)=\left\{\begin{array}{cc}
     1& \text{iff}\; u\!=\!v \\
    0 & \text{otherwise}.
\end{array}\right.
\end{equation}

Fig.\ref{fig:abs}d) shows the modularity graphs associated to the community label vectors in Fig.\ref{fig:abs}(c), referring to the different considered scales.

Our modularity graph representation has a double scope: firstly, it provides a robust multiscale framework for analyzing graph modularity unaffected by community numbering; and secondly, it offers an original view of the network structure, highlighting both the internal integrity of communities and the extent of their interaction with one another. By adopting this approach, we significantly enhance the interpretability of community structure analysis, paving the way for more accurate and insightful investigations into the functional organization of the brain.

\section{Application to Stochastic Block Models}
\label{sec:sbm}
In this section we apply our method on Stochastic Block Models (SBM) and we show how the modularity graph $\mathcal{G}^{\mathcal{M}\;(s) }$ describes the SBM data interactions.

With the aim to generate SBM graphs, we 
produce $K$ observations of a Stochastic Block Model known as  Planted Partition Model  (PPM). In this class ofngraph, the number of communities $P$ and the affinity matrix are fixed $\boldsymbol{\Gamma}$ fixed. Then, the  graph edges are randomly drawn with probability $p_{A_{ij}|\gamma_i,\gamma_j}(A_{ij}|\gamma_i,\gamma_j)$ distributed as in \eqref{eq:probmodel},  according to an assortative community structure.

\begin{figure}
    \centering{\includegraphics[scale=0.7]{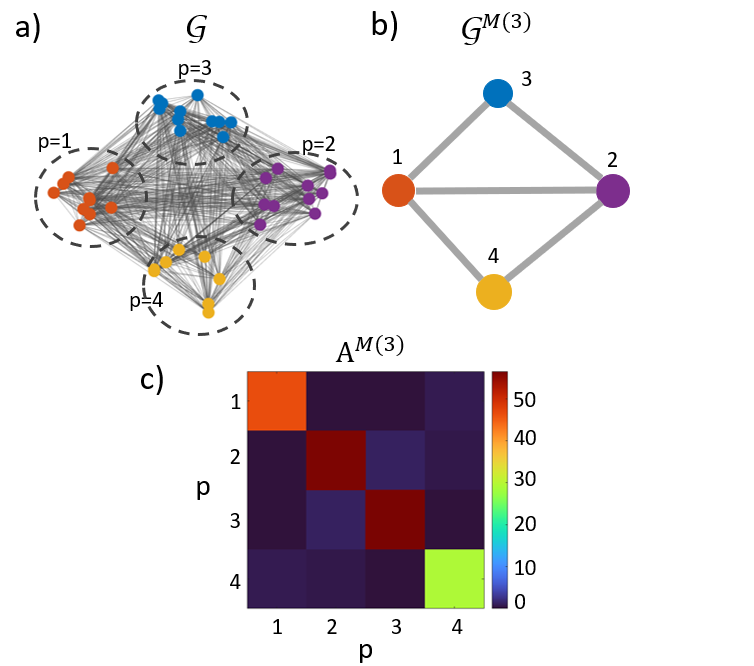}}
    \caption{Modularity graph $\mathcal{G}^{\mathcal{M}\;(s) }$ at $s=3$ with $P=4$ communities and $K=1$. In panel a) we have the original graph $\mathcal{G}$ with $N=30$, where each node is coloured according to its associated community label in $\gamma$ . In panel b) we have the modularity graph $\mathcal{G}^{\mathcal{M}\;(3)}$ with a number of nodes equal to $P=4$. In panel c) we have the corresponding $P\times P$ adjacency matrix $A^{\mathcal{M}\;(3)}$. }
    \label{fig:synthPP}
\end{figure}

 After generating the ground truth adjacency matrix $\mathbf{A}_{GT}$, we assume that it is observed in the presence of i.i.d. additive white Gaussian noise, with zero mean and variance $\sigma_w^2$ in the different trials $k\!=\!0,\cdots K\!-\!1$.  Thus, we obtain a trial dependent adjacency matrix $ \mathbf{A}^{(k)}$ as follows:

\begin{equation}    \mathbf{A}^{(k)} = \mathbf{A}_{GT}+ \mathbf{W}^{(k)}, k\!=\!0,...K\!-\!1
\label{eq:normerr}
\end{equation}

For the sake of concreteness, and without loss of generality, let us  consider the case with $N=30, P=4$ and $SNR=50$. Following the procedure presented in the previous section we obtain the multiscale modularity graph $\mathcal{G}^{\mathcal{M}\;(s) }$ and we show the corresponding adjacency matrix for $s=3$. The choice of the wavelet scale $s$ for the community detection appears natural in this case, since the number of graph communities $P$ was initially set, whereas in real data it must be estimated.

In Fig.\ref{fig:synthPP} we report the modularity graph $\mathcal{G}^{\mathcal{M}\;(s) }$ at $s=3$ that identifies $P=4$ communities. In panel a)  we have the graph representation, where the nodes are colored according to their community labeling. In panel b) we have the graphical representation of the modularity graph at the fixed scale. The modularity graph immediately shows which community has the largest weighted links,  which  communities  interact and how separated the network clusters are. In panel c)  we have the modularity graph adjacency matrix where we can identify different components. Indeed, the principal diagonal in $i,i$ contains the sum of the weights of the links belonging to the  $i$-th community while in the extradiagonal elements we have the sum of the weights of the links connecting  nodes across different clusters.

\section{Application to real EEG signals recorded during motor imagery}
\label{sec:eeg}
This section presents a proof-of-concept of the applicability of our framework to real EEG data acquired during motor imagery experiments and collected in a public dataset \cite{tangermann2012review}.  In this dataset, $7$ subjects are included, with real and synthetically generated. Here, we consider a real subject performing a binary motor imagery task, i.e. motor imagery of the left hand vs. rest with  $K=100$  trials per condition.

We resort to the similarity metric 
$\mathcal{S}\left(\mathbf{x}_i, \mathbf{x}_i\right)={\sum_{\omega \in \mathcal{B}}IC_{ij}(\omega)}/{\left|\left| {\mathcal{B}} \right|\right|}$ built considering the imaginary coherence $IC_{ij}(\omega)$ in the frequency band of interest $\mathcal{B}$ as in (\ref{eq:b}) and (\ref{eq:icoh}). Specifically, we select $\mathcal{B}$ in correspondence to the EEG \textit{beta} band $[12.5\!-\!30]$ Hz.

Among the possible alternatives for the choice of the appropriate functional connectivity (FC) \cite{gonzalez2021network}, we consider imagery coherence that has shown the ability to characterize motor imagery-based  brain computer interface (BCI) mental states \cite{cattai2021phase}. Imaginary coherence improves the FC estimation by filtering out instantaneous correlations, allowing a clearer description of how brain regions communicate during different tasks  \cite{nolte2004identifying}. 
In order to reduce the number of edges of the adjacency matrix we threshold it by keeping the $60\%$ of strongest connections.
We apply the community detection algorithm \cite{tremblay2014graph} that provides a multiscale community structure captured in $\gamma_i^{k,s}$ at the $k$-th trial and $s$-th scale. 

\begin{figure}
    \centering{\includegraphics[scale=0.75]{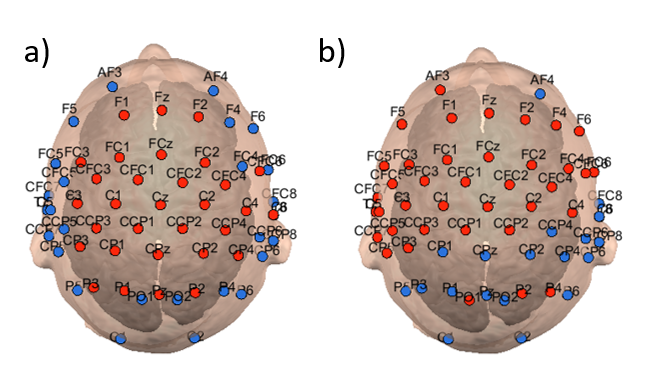}}
    \caption{Graphical representation of graph communities at $s=6$ and $k=1$ in $\beta$ frequency band during different motor imagery tasks.The colour associated to each node corresponds to the community label $\gamma^{1,6}$. In panel a) the task is motor imagery of the left hand, while panel b) reports community labels  at rest.}
    \label{fig:comreal}
\end{figure}

In Fig.\ref{fig:comreal} we have the representation of the community organization for $k=1$ at $s=6$, where $P=2$ communities are identified. In panel a) we report results associated with the motor imagery of the left hand while in panel b) we have communities detected in resting state. By the observation of one single trial, it is possible to remark the presence of a central community during the motor imagery of the left hand; this could be explained by the emergence of inter-hemispheric connections during that specific task \cite{fallani2013multiscale}.

Then, we derive graph $\mathcal{G}^{\mathcal{M}\;(s)}$ that correspond to the modularity graph at the scale $s$ by computing the  $s$-th scale adjacency matrix  $A^{\mathcal{M}\;(s)}_{pq}$ as in 
 \eqref{eq:AM}.
With the aim to measure the ability of those graphs to detect the correct mental state of the BCI user, we statistically compare the distribution of each element of the adjacency matrix of the modularity graphs across the experimental trials at a  fixed wavelet scale.

The discriminant power of modularity graph is quantified by the $t$-value, which measures the statistical separability  between two considered conditions (hypotheses); in this case, the two situations (hypotheses) correspond to the two mental states of i) motor imagery of the left hand and ii) rest.  Specifically, the $t$-value measures the difference, observed across different trials, of the estimated average values $A_{ij}^{M(s)}$, suitably normalized to compensate for the fluctuations due to the finite number of observations. 

\begin{figure}
    \centering{\includegraphics[scale=0.55]{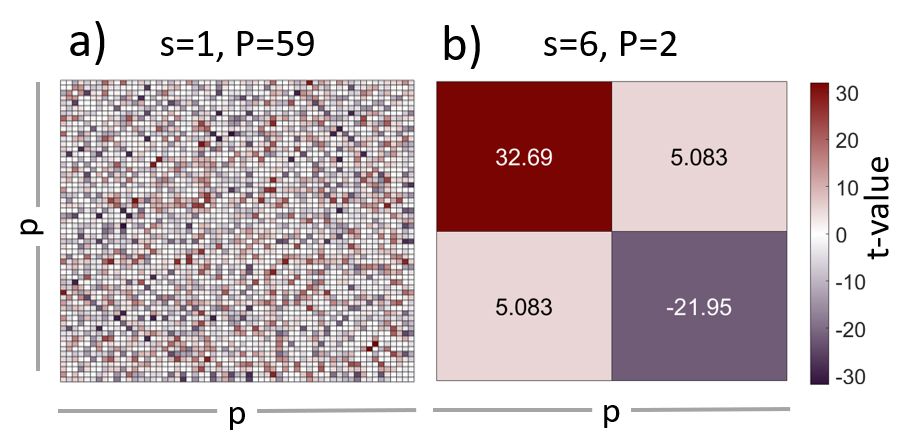}}
    \caption{Graphical representation of results of t-test between modularity graphs $\mathcal{G}^{\mathcal{M}\;(s) }$. In panel a) we have $s=1$ that corresponds to $P=59$, while in panel b) the case of  $s=6$ that corresponds to $P=2$ is reported. The colours represent  the t-values that are esplicitely written in panel b). All the results are significantly different between the two mental states, with $\rho=0.01$.}
    \label{fig:ttest}
\end{figure}

The results of the statistical tests are reported in terms of $t$-values,  with a significance value of $\rho=0.01$, meaning that the values $A_{pq}^{M(s)}$ observed under the two conditions are actually separable with probability $1\!-\!\rho=0.99$. In Fig.\ref{fig:ttest} we represent in a $P \times P$ matrix the t-values by highlighting the number of communities identified at each scale.  Specifically, we have in panel a) the case $s=1$, where the community detection algorithm derives $P=59$ communities, corresponding to the number of nodes $N$ in the graph. Increasing the wavelet scale for $s=2$ the number of detected communities remains $P=N=59$. At $s=3$ in the condition of motor imagery we have $P=3$ while during rest we have $P=59$, demonstrating a more structured organization in the brain to complete complex tasks. At $s=4,5,6$ the number of clusters corresponds to $P=2$ and in panel b) we report the results of the statistical analysis at $s=6$ in a $P\times P$ (i.e. $2 \times 2$) matrix. Our findings show that the adjacency matrices of the modularity graph change significantly during motor imagery, proving an original and useful tool to discriminate between human mental states during BCI tasks.

\subsection*{Remarks}
The multiscale community detection algorithm automatically identifies the number of communities in the two classes \cite{tremblay2014graph}. In this manner, the number of clusters can change in the mental states, preventing the statistically comparidon of the two associated adjacency matrices by simple t-tests. Other strategies can be envisaged, such as the selection of key communities based on a  scoring strategy, as their Jensen divergence under the two conditions \cite{cattai2021improving}. In order to extend and generalize the framework, a study on the percentage of links to be kept in the adjacency matrix is needed. Moreover, a tuning of fundamental parameters, as the frequency band \cite{cattai2021phase}, and a preprocessing step \cite{wierzgala2018most} are needed to effectively use those features in classifying mental states.


\section{Conclusions and further work}
\label{sec:concl}
 This paper introduced a novel graph-based feature, namely the modularity graph, representing the relationship across the graph communities. We applied the modularity graph to the brain functional connectivity network learned from real EEG data, and analyzed by means of a multiscale community mining algorithm. The modularity graph proved effective in quantifying the interactions between neuron clusters within the brain's network.   Statistical  analysis across multiple scales shows that  the modularity graph differs for the distinct   brain connectivity states associated with various motor imagery tasks, and it 
 paves the way  for future integration of the modularity graph within mental state classification systems.

\bibliography{main}
\bibliographystyle{ieeetr}

\end{document}